\documentclass[12pt,english]{article}
\pdfoutput=1
\usepackage[T1]{fontenc}
\usepackage{amssymb}
\usepackage{amsmath}
\usepackage{cite}
\usepackage{epsfig}
\usepackage[unicode=true,
 bookmarks=true,bookmarksnumbered=false,bookmarksopen=false,
 breaklinks=false,pdfborder={0 0 1},backref=false,colorlinks=true]
 {hyperref}
\usepackage{breakurl}
\usepackage[hang,flushmargin]{footmisc}
\usepackage{breakurl}
\usepackage{color}

\usepackage[latin1]{inputenc}
\usepackage{amsmath}
\usepackage{amsfonts}
\usepackage{amssymb}
\usepackage{graphicx}

\usepackage{pifont}

\makeatletter
\def\simgt{\mathrel{\lower2.5pt\vbox{\lineskip=0pt\baselineskip=0pt
           \hbox{$>$}\hbox{$\sim$}}}}
\def\simlt{\mathrel{\lower2.5pt\vbox{\lineskip=0pt\baselineskip=0pt
           \hbox{$<$}\hbox{$\sim$}}}}
\makeatother

\newcommand{\be}{\begin{equation}}
\newcommand{\ee}{\end{equation}}
\newcommand{\bea}{\begin{eqnarray}}
\newcommand{\eea}{\end{eqnarray}}
\newcommand{\eq}[2]{\be\begin{aligned}#1 \label{#2}\end{aligned}\ee}

\newcommand{\Eq}[1]{Eq.~\eqref{#1}}

\newcommand{\Sec}[1]{Sec.~\ref{#1}}

\newcommand{\App}[1]{App.~\ref{#1}}

\hypersetup{citecolor=blue,linkcolor=blue,urlcolor=blue}

\begin{document}

\interfootnotelinepenalty=10000
\baselineskip=18pt

\hfill CALT-TH-2020-060

\vspace{2cm}
\thispagestyle{empty}
\begin{center}
{\LARGE \bf
Symmetry and Unification from \\
\smallskip
 Soft Theorems and Unitarity\\
}
\bigskip\vspace{1cm}{
{\large Clifford Cheung and Zander Moss}
} \\[7mm]
 {\it ${}^a$Walter Burke Institute for Theoretical Physics\\[-1mm]
    California Institute of Technology, Pasadena, CA 91125 } \let\thefootnote\relax\footnote{\noindent e-mail: \url{clifford.cheung@caltech.edu}, \url{zander@caltech.edu}} \\
 \end{center}

\bigskip
\centerline{\large\bf Abstract}
\begin{quote} \small

We argue that symmetry and unification can emerge as byproducts of certain physical constraints on dynamical scattering.  To accomplish this we parameterize a general Lorentz invariant, four-dimensional theory of massless and massive scalar fields coupled via arbitrary local interactions.  Assuming perturbative unitarity and an Adler zero condition, we prove that any finite spectrum of massless and massive modes will necessarily unify at high energies into multiplets of a linearized symmetry. Certain generators of the symmetry algebra can be derived explicitly in terms of the spectrum and three-particle interactions.  Furthermore, our assumptions imply that the coset space is symmetric. 

\end{quote}

\setcounter{footnote}{0}

\newpage

\section{Introduction} \label{sec:introduction}

It is often said that symmetry is beautiful.  However, this view presumes that symmetry is a luxury with which a theory may or may not be blessed.  This intuition fails when symmetry is {\it required} by consistency and a less symmetric theory is simply impossible.  A classic example of this is gauge symmetry, which is not fundamental but rather mandated by more primitive principles such as Poincar\'e invariance and locality.  Indeed, it is well-known that by bootstrapping the scattering amplitudes of self-interacting massless vector particles directly from these underlying assumptions, one can derive all the properties of the gluon without the aid of gauge invariance (see  \cite{Benincasa:2007xk, Elvang:2015rqa, Cheung:2017pzi} and references therein).  In this approach notions such as charge conservation and the Lie algebra structure  emerge from these more basic concepts.

On the other hand, none of these arguments apply to global symmetries since they are not redundancies of description. So it is possible in principle for a theory to be imbued with more versus less global symmetry in a well-defined sense.    It then seems genuinely remarkable when a global symmetry, spontaneously broken at long distances, is miraculously and intricately restored at short distances.  From this perspective, theories are consecrated with unifying symmetries in the deep ultraviolet which are muddied at low energies due to the noninvariance of the vacuum.  Aspects of the emergent pion degrees of freedom, e.g.~their spectrum \cite{Nambu:1960tm,Goldstone:1961eq,Goldstone:1962es}, nonlinearly realized symmetries \cite{Coleman:1969sm,Callan:1969sn}, and soft behavior \cite{Adler:1964um,Adler:1965ga} are then taken to be derived properties of this underlying high-energy symmetry.

In this paper we pursue the exact opposite logic.  Ultraviolet unification and symmetry restoration are not assumed.  Instead, we will derive them as {\it consistency conditions} implied by an alternative set of physical principles naturally defined in the broken phase.   For our setup we study a general Lorentz invariant, four-dimensional theory of scalar fields with arbitrary masses and interactions.
Here the spectrum of the theory is defined to include {\it all} ultraviolet degrees of freedom and the interactions may be higher dimension or derivatively coupled.
We then impose four physical criteria:

\begin{itemize}
\item {\bf Perturbative Unitarity.} The dynamics are perturbative and unitary.

\item {\bf Locality.} The interactions are polynomials in derivatives.

\item {\bf Finite Spectrum.} The number of degrees of freedom is finite.

\item {\bf Soft Theorems.}   All amplitudes vanish in the soft limit of a massless mode.
\end{itemize} 

\noindent The first condition holds in any weakly coupled theory to which perturbation theory applies. 
 The second condition is a technical assumption imposed so that the tree amplitudes are rational functions of momenta.  It is violated if the theory has nonlocal interactions.  The third statement is required so that high-energy limits of amplitudes are well-defined.  The last and strongest assumption is a variation of the Adler zero condition \cite{Adler:1964um,Adler:1965ga} in which we demand that all tree-level amplitudes vanish when a massless degree of freedom is taken soft.

We will prove that these four conditions imply that the massless and massive degrees of freedom  necessarily unify into a multiplet which, in the high-energy limit, transforms {\it linearly} under a global symmetry.  The generators of the ultraviolet symmetry can actually be expressed explicitly in terms of the mass spectrum and three-particle interactions. 
 As one would expect, the ultraviolet theory need not be unique.  When our assumptions do not fix all parameters in the theory, those that remain simply label the allowed space of possible ultraviolet completions consistent with a given spectrum.  Interestingly, we find that our four criteria imply that the coset space of the symmetry breaking is symmetric.  Note that our approach differs from prior efforts on bootstrapping the massless sector alone, either from the Adler zero \cite{Low:2014nga, Cheung:2014dqa, Cheung:2015ota, Cheung:2016drk, Cheung:2018oki, Elvang:2018dco} without imposing unitarity, or from alternative ultraviolet considerations \cite{Carrasco:2019qwr}.

The very simplest theory satisfying the above four criteria---a theory of one massless and one massive particle---is easy to understand.  The only scalar potential consistent with our assumptions is the ``wine-bottle'' potential familiar from spontaneous symmetry breaking.  In particular, the form of the low-energy basin of the potential is dictated by the Adler zero while the shape of the rest of the potential is uniquely fixed by perturbative unitarity.   
 
The remainder of this paper is structured as follows.  In \Sec{sec:perturbative_unitarity}  we define a general theory of massless and massive scalars coupled through arbitrary local interactions.  We review how to extract the high-energy behavior of an amplitude using a minimal basis of kinematic invariants before generalizing this notion to off-shell operators.   We then present a proof that the assumptions of perturbative unitarity, locality, and a finite spectrum imply the existence of a field basis in which the Lagrangian takes a renormalizable form.  Afterwards, in  \Sec{sec:adler_symm} we impose an Adler zero condition on four- and five- particle scattering amplitudes, deriving a set of highly nontrivial constraints on the interactions of the theory.  These in turn imply the existence of unbroken and broken symmetry generators which act as linear and affine transformations that mix the massless and massive states. The broken generators are constructed explicitly from the cubic couplings and masses of the constrained Lagrangian, while their commutators produce a subalgebra of the unbroken generators, thus establishing that the coset space of broken symmetries is necessarily symmetric.
 It is then straightforward to show that in the high-energy limit these symmetries form a subgroup of special orthogonal rotations under which all fields transform linearly. In \Sec{sec:examples}, we demonstrate the connection between Adler zeros and symmetries explicitly through an example theory involving a single massive state and an arbitrary number of massless modes. We then present our conclusions and future directions in \Sec{sec:conclusions}.

\section{Perturbative Unitarity} \label{sec:perturbative_unitarity}

Consider a Lorentz invariant, four-dimensional theory of interacting scalar fields.  Here we define the spectrum so as to include all degrees of freedom in the theory, heavy or light.  Furthermore, we assume that these states are finite in number so it is possible to take a high-energy limit that exceeds all the physical mass thresholds in the theory.   Since unitarity forbids ghost modes even in the linearized theory, we are required to assume a quadratic dispersion relation, i.e.~$p^2 = m^2$.

At the nonlinear level we allow for an {\it a priori} arbitrary set of local interactions which are unbounded in the number of external fields.  However, we also assume that at any given number of external fields the interactions are at most polynomial in the external momenta\footnote{For simplicity we do not consider here terms involving the Levi-Civita tensor, although this would be an interesting avenue for future analysis.}.  The purpose of the latter condition is to forbid nonlocal interactions, which necessarily entail an infinite train of higher derivative corrections\footnote{For this reason our conclusions will not be applicable to any theory of extended objects such as string theory.}.  While nonlocalities generically arise when integrating out fields, this is not permitted here since our spectrum is defined to include all heavy and light degrees of freedom.  No particles have been implicitly integrated out.   

\subsection{Amplitudes at High Energy} \label{sec:he_behavior}

An $N$-particle scattering amplitude $\mathcal{A}_N$ is a function of the momenta $\{ p_a \}$ and flavor indices $\{ I_a \}$ of the external legs, where $1 \leq a \leq  N$.   Naively, the high-energy behavior of $\mathcal{A}_N$ is obtained by rescaling $p_a \rightarrow z p_a$ for large $z$.  However, this operation is not self-consistent because it does not preserve the on-shell condition for massive particles, $p_a^2 = m_{I_a}^2$, and fails to account for the fact that certain combinations of momentum, e.g. $\sum_{a=1}^N p_a =0$, do not actually scale as $z$.

Nevertheless it is trivial to extract the high-energy behavior on-shell, provided we first reduce to a minimal basis of kinematic invariants (see Appendix B of \cite{Cheung:2017pzi}).  \textit{A priori}, $\mathcal{A}_N$ depends on $N(N+1)/2$ invariants of the form $p_a \cdot  p_b$ for $1\leq a,b \leq N$.  We can enforce total momentum conservation by substituting all invariants involving one of the momenta in terms of the others, leaving just $N(N-1)/2$.  The on-shell conditions then impose $N$ linear relations among the remaining invariants, giving a final tally of $N(N-3)/2$ independent objects.  For our explicit calculations we choose for this ``minimal kinematic basis'' the set\footnote{Complications involving four-dimensional Gram determinant identities can be ignored since we are interested in at most five-particle scattering, which still depends on four linearly independent momentum vectors even after accounting for momentum conservation.  }
\eq{
\{ \, p_a  \cdot p_b \, \} \quad \textrm{where} \quad 1 \leq a \leq N-3 \textrm{ and } a< b \leq N-1 .
}{eq:min_basis}
Note that we have eliminated $p_{N-2}\cdot p_{N-1}$ using the on-shell condition $p_N^2 = m_{I_N}^2$.

To take the high-energy limit, we transform the minimal kinematic basis by
\eq{
p_a \cdot p_b \rightarrow z^2 \, p_a \cdot p_b,
}{eq:HE_limit}
for large $z$, which probes high energies without leaving the on-shell surface.  As discussed in \App{app:PU},  the assumptions of perturbative unitarity, locality, finite mass spectrum, and Adler zeros for massless particles in four spacetime dimensions imply that the $N$-particle tree-level scattering amplitudes must satisfy the scaling bound\cite{Cornwall:1974km}
\eq{
\mathcal{A}_N (z\rightarrow \infty) \lesssim z^{4-N}.
}{eq:scaling_bound}
More precisely, the modulus of any $N$-particle scattering amplitude is strictly bounded at large $z$ by $|\mathcal{A}_N| \leq q |z|^r$ for some $q>0$ and $r \leq 4-N$. 
As we will see shortly, this condition strongly constrains the allowed form of the Lagrangian.

Note that this inequality is exactly saturated in any Lorentz invariant four-dimensional quantum field theory with dimensionless couplings, e.g. Yang-Mills theory or scalar $\phi^4$ theory.   Furthermore, we emphasize that \Eq{eq:scaling_bound} is a necessary but not sufficient condition for perturbative unitarity.  For instance, tree-level amplitudes in quantum electrodynamics satisfy \Eq{eq:scaling_bound} but will violate perturbative unitarity at energies near the Landau pole. 

\subsection{Lagrangians at High Energy}

High-energy behavior is even harder to discern at the level of the Lagrangian.  As a simple example, consider the cubic operator $ \phi \,\partial^\mu \phi \, \partial_\mu \phi $, which naively implies $\mathcal{A}_3 \sim {\cal O}(z^2)$ scaling for three-particle scattering.  Instead, $\mathcal{A}_3 \sim {\cal O}(z^0)$ because all the invariants $p_a \cdot p_b$ can expressed in terms of external masses, i.e.~the minimal kinematic basis is empty.  As another example, the quartic operator $ \phi^2  \partial^\mu \phi \, \partial_\mu \phi$ naively implies $\mathcal{A}_4 \sim {\cal O}(z^2)$ but instead $\mathcal{A}_4 \sim s+t+u \sim {\cal O}(z^0)$ due to momentum conservation, on-shell conditions, and Bose symmetry.   

To circumvent this annoyance we make high-energy scaling manifest at the level of the Lagrangian by defining a ``minimal operator basis'' in analogy with the the minimal kinematic basis for amplitudes.  Consider a general off-shell $N$-particle operator, $F_N(\partial_1, \partial_2, \cdots , \partial_N) \phi_{I_1} \phi_{I_2} \cdots \phi_{I_N}$, where $\partial_a$  denotes a derivative acting on the field $\phi_{I_a}$ with $1\leq a \leq N$.  Since we have assumed that all interactions are polynomially bounded in derivatives, $F_N$ is a polynomial in its arguments.

If a subset of the fields are the same flavor, and thus indistinguishable, we explicitly symmetrize $F_N$ on their corresponding labels.  In parallel with the amplitudes approach, $F_N$ depends \textit{a priori} on $N(N+1)/2$ invariants $\partial_a \partial_b$ for $1 \leq a,b \leq N$.  To derive an analog of the minimal kinematic basis for amplitudes in \Eq{eq:min_basis}, we use integration by parts to shuffle all derivatives acting on $\phi_{I_N}$ onto the other fields, thus eliminating $\partial_N$ by the momentum conservation constraint.   By performing a field redefinition we can effectively set $\partial_a^2 = -m_{I_a}^2$ for $1\leq a \leq N-1$ at the level of the action, modulo contact terms which have more than $N$ fields and can thus be absorbed into the definition of higher order operators\footnote{Concretely, the field transformation $\phi \rightarrow \phi + \delta \phi$ will induce the variation of the action, $S\rightarrow S + \delta S$ where $\delta S = -\int \delta \phi [(\Box + m^2) \phi + \cdots]$.  Here $\delta \phi$ starts at quadratic order in fields and the ellipses denote higher order terms.  By an appropriate choice of $\delta \phi$, any term in the action proportional to $\Box \phi$ can be substituted for $-m^2 \phi$.}.  Here we also eliminate $\partial_{N-2}\partial_{N-1}$, which via integration by parts can be related to $\partial_N^2 = -m_{I_N}^2$.

It is then mechanical to construct the minimal operator basis, first starting with operators with the fewest number of fields, and then working our way up.  In the case of $F_3$, all derivatives of fields can be converted, via integration by parts, to D'Alembertian operators acting on products of the other fields. Each D'Alembertian can then be eliminated in favor of an $m^2$ factor using a field redefinition.  For example, an operator like $\partial^\mu \phi_1 \partial_\mu  \phi_2 \, \phi_3$ is equivalent upon integration by parts to $(\phi_1\phi_2 \Box \phi_3-\Box \phi_1\phi_2 \phi_3-\phi_1 \Box\phi_2 \phi_3)/2$, which on equations of motion is equivalent to the bare potential term $\phi_1 \phi_2 \phi_3$ modulo terms quartic or higher in the field.  So in the minimal operator basis,  $\phi_1 \phi_2 \phi_3$ is the only allowed object at cubic order in the fields. As advertised, it exactly manifests the correct ${\cal O}(z^0)$ high-energy scaling required of any three-particle amplitude.  We proceed to apply the same procedure to $F_4$, $F_5$, etc., yielding a Lagrangian for which all operators correctly manifest their high-energy behavior.

\subsection{Deriving Renormalizability}

We have shown that in the minimal operator basis, the only possible cubic operator is a potential term.   Hence, {\it any} scalar Lagrangian can be put into the form\footnote{Throughout, all repeated indices are taken to be summed over unless otherwise specified.}
\begin{align} \label{eq:min_op_lag}
\mathcal{L} = -\frac{1}{2}\partial^\mu \phi_I \partial_\mu \phi_I - \frac{1}{2} m_I^2 \phi_I \phi_I -\frac{1}{3!} \alpha_{IJK} \phi_I \phi_J \phi_K +\cdots,
\end{align}
where the ellipses denote all possible interactions at quartic order in fields or higher and involving an arbitrary but finite number of derivatives.

To study the theory at quartic order in the fields we analyze the four-particle scattering amplitude computed from the Feynman diagrams corresponding to the Lagrangian in \Eq{eq:min_op_lag}.  In the high-energy limit we obtain $\mathcal{A}_4(z) = \mathcal{A}_4^{\rm cont}(z) +  \mathcal{A}_4^{\rm fact}(z)$, where we have distinguished between ``contact'' contributions from the quartic vertex and the $s$-, $t$-, and $u$-channel ``factorization'' diagrams coming from the cubic vertices.  Since the cubic operators are derivative-free, the factorization term is simply $\mathcal{A}_4^{\rm fact}(z) \sim {\cal O}(z^{-2})$.  From the perturbative unitarity bound in \Eq{eq:scaling_bound} we know that $\mathcal{A}_4(z)\lesssim {\cal O}(z^{0})$, which then implies that $\mathcal{A}_4^{\rm cont}(z) \lesssim {\cal O}(z^{0})$.  Since we have assumed that all interactions are polynomials in the momenta, we deduce that the quartic vertex is a constant.

The same logic applies at quintic order in the fields.  Splitting the amplitude as $\mathcal{A}_5(z) = \mathcal{A}_5^{\rm cont}(z) +  \mathcal{A}_5^{\rm fact}(z)$, we see that the contribution from all factorization diagrams must scale \textit{at most} as $\mathcal{A}_5^{\rm fact}(z) \sim {\cal O}(z^{-2})$ since the cubic and quartic vertices are both constants. Perturbative unitarity then implies that $\mathcal{A}_5(z) \lesssim {\cal O}(z^{-1})$, which, because of our assumption of locality is impossible unless $\mathcal{A}_5^{\rm cont}(z) = 0$, so the quintic contact term must vanish. Since we are working in a minimal operator basis it then follows that the quintic Lagrangian operator must vanish completely.   At sextic order and higher, the scaling bound is increasingly negative and the same reasoning applies iteratively.  Hence, all Lagrangian operators are absent except for those at quartic or lower orders.

In summary, we have shown that under the assumptions of perturbative unitarity, locality, and a finite mass spectrum, the Lagrangian for any Lorentz invariant, four-dimensional theory of scalars is equivalent---up to a field redefinition that leaves all scattering amplitudes invariant---to the well-known renormalizable form
\eq{
\mathcal{L} = -\frac{1}{2}\partial^\mu \phi_I \partial_\mu \phi_I - \frac{1}{2} m_I^2 \phi_I \phi_I -\frac{1}{3!} \alpha_{IJK} \phi_I \phi_J \phi_K -\frac{1}{4!} \beta_{IJKL} \phi_I \phi_J \phi_K \phi_L ,
}{eq:L_renormalizable}
where $\alpha_{IJK}$ and $\beta_{IJKL}$ are symmetric on all indices.
 This argument rules out the possibility of a perturbatively unitary theory with derivatively coupled interactions, so we learn e.g.~that there is no variation of the nonlinear sigma model that is well-behaved at high energies without additional modes.
 Here we emphasize again that our argument relies crucially on the assumption of polynomial boundedness.    

\section{Soft Theorems and Unification} \label{sec:adler_symm}


We are now ready to derive explicit constraints on the Lagrangian in \Eq{eq:L_renormalizable} derived from the Adler zero conditions.    As we will see, the resulting constraints will directly imply the existence of a symmetry connecting the massless and massive degrees of freedom.  Past work has applied similar logic to the pion sector alone \cite{Susskind:1970gf,Low:2014nga} as well to more exotic theories such as Dirac-Born-Infeld theory, the Galileon \cite{Cheung:2014dqa, Cheung:2015ota, Cheung:2016drk, Cheung:2018oki}, and more recently even the Navier-Stokes equation \cite{Cheung:2020djz}.

\subsection{Adler Zero Constraints}

As discussed earlier, we assume an Adler zero condition that mandates the vanishing of all scattering amplitudes in the limit where any massless particle is taken soft.  For simplicity we refer to all massless fields as pions and all massive fields as sigmas.  

Without loss of generality, consider an $N$-particle tree-level scattering amplitude in which leg 1 is a soft pion, so $p_1 \rightarrow z p_1$ for small $z$. Note that here it is crucial that the minimal kinematic basis defined in \Eq{eq:min_basis} eliminates redundant kinematic invariants involving legs $N-1$ and $N$ but does not interfere with leg 1.  In the minimal kinematic basis, this soft scaling sends
\eq{
p_1  \cdot p_a \rightarrow z \,  p_1  \cdot p_a ,
}{eq:soft}
for all $a$.   The soft-deformed amplitude has a series expansion $\mathcal{A}_N(z) \sim z^{-1} +z^{0} +\cdots$, where the leading ${\cal O}(z^{-1})$ term is a soft pole and the subleading ${\cal O}(z^0)$ term is regular.  The Adler zero condition implies that the $N$-particle tree-level scattering amplitude satisfies
\eq{
\mathcal{A}_N (z\rightarrow 0) = 0, 
}{}
so the leading and subleading terms are both zero.  Moreover, the coefficients of these terms are themselves complicated rational functions of the minimal kinematic basis of invariants, so each Adler zero condition actually dictates multiple constraints on the couplings of the theory.   

In what follows, we compute the Adler zero constraints for four- and five-particle tree-level scattering amplitudes.  In principle, additional constraints can arise at six-particle scattering and higher but they will not be necessary for the arguments in this paper.

\subsubsection{Four-Particle Amplitude}

Using the Feynman diagrams defined from the Lagrangian in \Eq{eq:L_renormalizable}, we compute the soft-deformed four-particle scattering amplitude,
\begin{align}
\mathcal{A}_4(z) = &- \sum_N \frac{\alpha_{I J N}\alpha_{N KL}}{(m^2_{J} - m^2_{N}) + 2z(p_1\cdot p_2)} 
-\sum_N \frac{\alpha_{I K N}\alpha_{N JL}}{(m^2_{K} - m^2_{N}) + 2z(p_1\cdot p_3)}\\ \nonumber
&-\sum_N \frac{\alpha_{I L N}\alpha_{N JK}}{(m^2_{L} - m^2_{N}) - 2z(p_1\cdot p_2 + p_1\cdot p_3)}
-\beta_{IJKL},
\end{align}
where $I$ is a pion flavor index, $J,K,L$ are flavor indices of any type, and $N$ runs over all possible flavors of the exchanged scalar.  We then take $z\rightarrow 0$ to extract the soft limit.

The Adler zero implies that the leading ${\cal O}(z^{-1})$ soft pole in $\mathcal{A}_4(z)$ is zero, so
\begin{align}
-\sum_{N | m^2_N = m^2_{J}}\frac{\alpha_{IJ N}\alpha_{N KL}}{2z(p_1\cdot p_2)} -\sum_{N | m^2_N = m^2_{K}}\frac{\alpha_{IK N}\alpha_{N JL}}{2z(p_1\cdot p_3)} +\sum_{N | m^2_N = m^2_{L}}\frac{\alpha_{IL N}\alpha_{N JK}}{2z(p_1\cdot p_2 + p_1\cdot p_3)}  =0,
\end{align}
where the sums are restricted to the subset of intermediate states which are mass-degenerate with an external state.
The above expression must vanish for all possible kinematics.  We can multiply through by a common denominator so that the left-hand side becomes a polynomial in $(p_1\cdot p_2)^2$, $(p_1\cdot p_2) (p_1\cdot p_3)$, and $(p_1\cdot p_3)^2$. Since these are all independent kinematic invariants, their coefficients separately vanish, and are in fact equivalent under permutations to
\begin{align}
\sum_{N | m^2_N = m^2_{J}}\alpha_{IJ N}\alpha_{N  KL} = 0.
\end{align}
For the choice  $ I = L$ and $J = K$, this reduces to the condition
\begin{align}
\sum_{N | m^2_N = m^2_{J}}\left(\alpha_{IJ N}\right)^2 = 0,
\end{align}
so each term in the sum vanishes independently because the coupling constants are real.   Recalling that $I$ indexes a massless pion, we relabel indices to obtain our final condition,
\begin{align} \label{eq:adler4_singular}
\alpha_{IJK}=0 \quad \textrm{for} \quad m^2_I=0 \quad \textrm{and} \quad m^2_{J}=m^2_K.
\end{align}
Said another way, in this field basis the Adler zero {\it forbids} any cubic interaction between a pion and any two particles of the same mass.  In hindsight this is obvious because the ${\cal O}(z^{-1})$ soft pole arises when the exchanged particle is mass-degenerate with an external state.

Meanwhile, the subleading ${\cal O}(z^{0})$ term in $\mathcal{A}_4(z)$ also vanishes,
\begin{align}\label{eq:adler4_regular}
\sum_{N | m^2_N \neq m^2_{J}}\frac{\alpha_{IJ N}\alpha_{N KL}}{m^2_{J} - m^2_N} +\sum_{N | m^2_N \neq m^2_{K}}\frac{\alpha_{IK N}\alpha_{N JL}}{m^2_{K} - m^2_N} +\sum_{N | m^2_N \neq m^2_{L}}\frac{\alpha_{IL  N}\alpha_{N JK}}{m^2_{L} - m^2_N} + \beta_{IJKL} = 0,
\end{align}
yielding another nontrivial constraint on the couplings and masses of the theory.

\subsubsection{Five-Particle Amplitude}

The calculation of the soft-deformed five-particle amplitude $\mathcal{A}_5(z)$ is straightforward but the resulting expression is quite complicated so we will not display it here.   Again we observe ${\cal O}(z^{-1})$ soft poles which arise when exchanged states are mass-degenerate with external states, so these terms vanish on the the condition of \Eq{eq:adler4_singular}.
Meanwhile, the Adler zero condition enforces the vanishing of the ${\cal O}(z^0)$ term of $\mathcal{A}_5(z)$, 
\begin{align}\label{eq:adler5_regular}
&\sum_{N | m^2_{N} \neq m^2_{J}} \frac{\alpha_{IJ N }\beta_{N KLM}}{m^2_{J} - m^2_{N}} + \sum_{N | m^2_{N} \neq m^2_{K}} \frac{\alpha_{IK N }\beta_{N JLM}}{m^2_{K} - m^2_{N}}\\ \nonumber
+& \sum_{N | m^2_{N} \neq m^2_{L}} \frac{\alpha_{IL N }\beta_{N JK M}}{m^2_{L} - m^2_{N}}+ 
\sum_{N | m^2_{N} \neq m^2_{M}} \frac{\alpha_{IM N }\beta_{N J K L}}{m^2_{M} - m^2_{N}} = 0,
\end{align}
where again $I$ is a pion flavor index and $J, K, L, M$ are flavor indices of any type.
Here we have presented a simplified version of the constraint by taking the limit $p_2\cdot p_3, p_2 \cdot p_4 \rightarrow \infty$, which is permitted since these are independent kinematic invariants and the Adler zero applies to any on-shell configuration.

\subsection{Symmetry Constraints}

Before determining the relationship between soft theorems and symmetry it will be helpful to know beforehand what to look for---that is, the mechanical sense in which spontaneously broken global symmetries constrain couplings and masses in the broken phase.    With these constraints in hand we will then show how they coincide exactly with those derived from the Adler zero.

\subsubsection{Spontaneous Symmetry Breaking}

Typically it is assumed that in the unbroken phase there is a multiplet $\Phi_I$ transforming in a linear representation of the symmetry group $G$, so
\begin{align}
\Phi_I \rightarrow \Phi_I +  W_{IJ} \Phi_J,
\end{align}
where $W_{IJ}$ is a generator of $G$.  However, in the broken phase the fields  acquire vacuum expectation values $v_I = \langle \Phi_I\rangle$ which are invariant only under a subgroup $H \subset G$.  Expanding about true vacuum, $ \Phi_I  = v_I + \phi_I $, we find that the field fluctuations $\phi_I$ transform as
\begin{align}
\phi_I \rightarrow \phi_I + W_{IJ}{\phi}_J +  W_{IJ} v_J,
\end{align}
which is the composition of a linear transformation and a constant shift.  

All transformations are classified according to whether or not they leave the vacuum invariant.  If $W_{IJ} v_J =0$ then the transformation is linear and corresponds to an unbroken generator in $H$ that we denote by ${\cal T}$.  If $W_{IJ} v_J\neq0$ then the transformation is affine, i.e.~is a composition of both a linear and shift component, and corresponds to a broken generator in $G/H$ that we denote by ${\cal X}$.   The unbroken and broken generators ${\cal T}$ and ${\cal X}$ act on the fields as
\eq{
\phi_I \rightarrow  \phi_I + {\cal T} \phi_I \qquad  \textrm{for} \qquad  {\cal T} \phi_I  &=  T_{IJ}\phi_J\\
\phi_I \rightarrow  \phi_I + {\cal X} \phi_I \qquad \textrm{for} \qquad {\cal X} \phi_I  &=X_{IJ}\phi_J + \lambda_I,
}{eq:affine_trans}
where $T_{IJ}v_J=0$ and $X_{IJ} v_J \neq 0$, and the shift is $\lambda_I = X_{IJ}v_J$.  
So the unbroken generators ${\cal T}$ are realized as linear transformations while the broken generators ${\cal X}$ are realized as affine transformations.
On occasion it will be convenient to express the unbroken and broken generators explicitly in terms of the linear and shift components of their corresponding transformations, so we will sometimes write ${\cal T } = (T_{IJ}, 0)$ and ${\cal X } = (X_{IJ}, \lambda_I)$.  

\subsubsection{Linear Symmetries}

Under the linear symmetry in \Eq{eq:affine_trans}, the Lagrangian in \Eq{eq:L_renormalizable} transforms as
\begin{align} \label{eq:delta_l_linear}
\mathcal{L} &\rightarrow \mathcal{L} - T_{IJ}(\partial^\mu \phi_I \partial_\mu \phi_J + m_I^2 \phi_I \phi_J)-\frac{1}{2}\alpha_{IJN}T_{NK} \phi_I \phi_J \phi_K -\frac{1}{3!}\beta_{IJKN}T_{NK}\phi_I \phi_J \phi_K \phi_L .\end{align}
For this to be a symmetry of the action, all additional terms on the right-hand side must vanish, implying a set of nontrivial constraints,
\begin{gather} \label{eq:linear_antisymm}
T_{IJ} +T_{JI}=0 \\ \label{eq:linear_2}
T_{IJ}m^2_I + T_{JI} m^2_J = 0\\ \label{eq:linear_3}
\sum_N \left( \alpha_{IJN}T_{NK} + \alpha_{IKN}T_{NJ} + \alpha_{JKN}T_{NI} \right)=0\\ \label{eq:linear_4}
\sum_N \left( \beta_{IJKN}T_{NL} + \beta_{IJLN}T_{NK} + \beta_{IKLN}T_{NJ} + \beta_{JKLN}T_{NI} \right)= 0.
\end{gather}
Here, repeated indices are {\it not} summed unless accompanied by an explicit summation, and we have manifestly symmetrized on indices due to Bose symmetry.  

\Eq{eq:linear_antisymm} implies that $T_{IJ}$ is antisymmetric, which is obvious because any linear symmetry of the Lagrangian in \Eq{eq:L_renormalizable} has no explicit momentum dependence and thus must be a symmetry of the kinetic term and potential term independently.  Since the kinetic term $\partial^\mu \phi_I \partial_\mu \phi_I$ is invariant under special orthogonal rotations, $T_{IJ}$ must be antisymmetric.  
Meanwhile, \Eq{eq:linear_2} implies that $T_{IJ}$ only acts nontrivially on states with equal mass, $m_I= m_J$. This is again obvious since an unbroken symmetry generator should only act on a subspace of mass-degenerate states.

\subsubsection{Affine Symmetries} Under the affine symmetry defined in \Eq{eq:affine_trans}, the Lagrangian in \Eq{eq:L_renormalizable} transforms as
\eq{
\mathcal{L} &\rightarrow  \mathcal{L} - m^2_I\lambda_I \phi_I -  X_{IJ}\partial^\mu \phi_I \partial_\mu \phi_J  -(m_I^2 X_{IJ} + \frac{1}{2}\alpha_{IJK}\lambda_K)\phi_I \phi_J\\ &\qquad-\frac{1}{2}(\alpha_{IJN}X_{NK} + \frac{1}{3}\beta_{IJKL}\lambda_L)\phi_I \phi_J \phi_K -\frac{1}{3!}\beta_{IJKN}X_{NK}\phi_I \phi_J \phi_K \phi_L ,}{eq:delta_l}
implying the following constraints,
\begin{gather} \label{eq:affine_1}
m^2_I \lambda_I = 0\\ 
\label{eq:delta_antisymm}
X_{IJ} +X_{JI}=0 \\
\label{eq:affine_2}
X_{IJ}m^2_I + X_{JI} m^2_J + \sum_K\alpha_{IJK}\lambda_K=0\\ \label{eq:affine_3}
\sum_N\left(\alpha_{IJN}X_{NK} + \alpha_{IKN}X_{NJ} + \alpha_{JKN}X_{NI}\right) + \sum_L\beta_{IJKL}\lambda_L=0\\ \label{eq:affine_4}
\sum_N\left(\beta_{IJKN}X_{NL} + \beta_{IJLN}X_{NK} + \beta_{IKLN}X_{NJ} + \beta_{JKLN}X_{NI}\right) = 0,
\end{gather}
where, once again, repeated indices are {\it not} summed unless accompanied by an explicit summation.
\Eq{eq:affine_1} implies that the components of the vector $\lambda_I$ are nonzero only when $m_I =0$.  Thus $\lambda_I$ only has support on the subspace of massless pion fields.  This is of course required of any nonlinearly realized symmetry.   Also, as before \Eq{eq:delta_antisymm} implies that $X_{IJ}$ is an antisymmetric generator of the special orthogonal group.  

We emphasize that the affine constraints in Eqs.~(\ref{eq:affine_1})-(\ref{eq:affine_4}) are satisfied by $X_{IJ}$ even after shifting arbitrarily by any $T_{IJ}$ which happens to satisfy the linear constraints in Eqs.~(\ref{eq:linear_antisymm})-(\ref{eq:linear_4}).   This is expected because ${\cal X} = (X_{IJ},\lambda_I)$ is a generator of the coset space $G/H$ and is thus only defined modulo the addition of a generator ${\cal T} = (T_{IJ} ,0)$ of the unbroken group $H$.

Finally, let us note that it is straightforward but tedious to prove closure of the symmetry algebra.  In particular, taking any combination of commutators of unbroken generators ${\cal T}$ satisfying Eqs.~(\ref{eq:linear_antisymm})-(\ref{eq:linear_4}) or broken generators ${\cal X}$ satisfying Eqs.~(\ref{eq:affine_1})-(\ref{eq:affine_4}), we obtain new generators that also satisfy these constraints.

\subsection{Symmetry from Soft Theorems}

At last, we are equipped to demonstrate how symmetry emerges from soft theorems and perturbative unitarity. Here we make no direct reference to vacuum expectation values or spontaneous symmetry breaking.  Instead, we simply show that the Adler zero conditions derived in Eqs.~(\ref{eq:adler4_singular})-(\ref{eq:adler5_regular}) imply the existence of affine transformations that {\it precisely} satisfy the conditions in Eqs.~(\ref{eq:affine_1})-(\ref{eq:affine_4}) required of any generator of a spontaneously broken symmetry.   
In fact, we will be able to constructively derive explicit formulas for \textit{all} the broken generators ${\cal X} = (X_{IJ},\lambda_I)$ and for a subalgebra of unbroken generators ${\cal T} = (T_{IJ},0)$.
 We will also learn that the coset space $G/H$ is symmetric. Lastly, we comment on the restoration of these symmetries at high energy, establishing unification as a consequence of unitarity and Adler zeros.

\subsubsection{Constructing Generators}

First, consider the Adler zero condition for a pion labelled by the flavor vector $\lambda_I$.  Since these states are massless, we have that $m_I=0$ and so \Eq{eq:affine_1} holds trivially.
Second, let us {\it define} the broken generators to be ${\cal X} =(X_{IJ},\lambda_I)$ such that
\begin{align} 
\label{eq:affine_def}
X_{IJ} =
\left\{ \begin{array}{cc}
- \dfrac{1}{m^2_I - m^2_J} \sum_K \alpha_{IJK} \lambda_K \qquad &, \qquad  m^2_I \neq m^2_J  \\ \\
0 \qquad &, \qquad  m^2_I = m^2_J  
\end{array}
\right\} .
\end{align}
Here $X_{IJ}$ is manifestly antisymmetric, thus satisfying \Eq{eq:delta_antisymm}.  Furthermore, by construction $X_{IJ}$ satisfies \Eq{eq:affine_2} for $m_I^2 \neq m_J^2$.   Third, \Eq{eq:adler4_singular} trivially implies \Eq{eq:affine_2} for $m_I^2 = m_J^2$.
Fourth, by contracting the free pion index in \Eq{eq:adler4_regular} and (\ref{eq:adler5_regular}) with $\lambda_I$ we immediately obtain \Eq{eq:affine_3} and (\ref{eq:affine_4}).    Thus, we have shown that $X_{IJ}$ defined in \Eq{eq:affine_def} satisfies all the requirements expected of a broken symmetry generator.  
For later convenience, let us also define 
\eq{
X_{IJ} &= \sum_K X_{IJ,K} \lambda_K,
}{eq:Xdef} 
where the index after the comma in $X_{IJ,K}$ is implicitly projected down to the pion subspace since that is where $\lambda_I$ has nonzero support.  On the other hand, the indices before the comma are general and can have support on both the pion and sigma subspaces.  Thus, each pion field direction maps to some broken generator  $X_{IJ,K}$.

It is now possible to derive formulas for some of the unbroken symmetry generators. Consider an amplitude for which legs 1 and 2 are pion fields with flavor indices $I,J$ and legs 3 and 4 have arbitrary flavor indices $K,L$. In this instance we have the choice of taking the soft limit of either leg 1 or leg 2. This pair of soft limits corresponds to the four-particle ${\cal O}(z^0)$ Adler zero constraint from \Eq{eq:adler4_regular}, together with the same condition with $I$ and $J$ swapped. The difference of these equations yields a new constraint on a commutator \cite{ArkaniHamed:2008gz},
\begin{align} \label{eq:adler4_commutator}
(m^2_{K} - m^2_{L})\sum_N\left(X_{K N ,I}X_{N L,J} - X_{K N ,J}X_{N L,I}\right) = 0,
\end{align}
with indices not implicitly summed.  Note that even though legs 1 and 2 are bosons, the difference between constraints is not trivially zero since each constraint is derived from a different kinematic region in which either leg 1 or leg 2 is soft. 

Next, we contract the commutator appearing in this constraint with an arbitrary antisymmetric tensor $\lambda_{IJ}$ which only has support on the subspace of pions.  Relabeling indices, we then obtain an expression for an unbroken generator ${\cal T} =(T_{IJ},0)$ where
\eq{
T_{IJ} = \sum_{K,L} T_{IJ, KL} \lambda_{KL} \qquad \textrm{where} \qquad   T_{IJ, KL} = X_{IN,K} X_{NJ,L}-X_{JN,K} X_{NI,L}.
}{eq:Tdef}
As before, the indices after the comma are implicitly projected down to the pion subspace and the indices before the comma are general. Note that in general, the $T_{IJ}$ constructed from \Eq{eq:Tdef} need not span the full set of unbroken generators. From this definition of $T_{IJ}$ and \Eq{eq:adler4_commutator} we see that \Eq{eq:linear_antisymm} and \Eq{eq:linear_2} are automatically satisfied.  This implies that $T_{IJ}$ is a generator of the special orthogonal group which only connects fields of equal mass.   Last but not least, by inserting \Eq{eq:Tdef} into \Eq{eq:linear_3} and \Eq{eq:linear_4}, we obtain  \Eq{eq:adler4_regular} and \Eq{eq:adler5_regular} after a bit of algebra, verifying that $T_{IJ}$ is an unbroken symmetry generator.  

\subsubsection{Structure of the Symmetry Algebra}

Armed with explicit formulas for certain symmetry generators in \Eq{eq:affine_def} and \Eq{eq:Tdef}, we are now able to deduce some interesting facts about the symmetry algebra.

First, we have actually rederived a version of Goldstone's theorem \cite{Goldstone:1961eq} which says that there is a bijective mapping from the space of pions to the space of broken generators $(X_{IJ},\lambda_I)$.  Injectivity holds because the Adler zero constraint is parameterized by the flavor of the soft pion, $\lambda_I$, which is both the shift vector $\lambda_I$ and the vector used to construct $X_{IJ}$ via \Eq{eq:affine_def}. Any pair of distinct pion states gives rise to distinct shifts, and therefore distinct broken generators.  To establish surjectivity, observe that the number of independent shift vectors $\lambda_I$ cannot exceed the dimension of the space of pions---otherwise one of these vectors would have to act on a massive mode, which is forbidden by \Eq{eq:affine_1}.  Moreover, even if a pair of broken generators ${\cal X} = (X_{IJ} , \lambda_I)$ and  ${\cal X}' = (X_{IJ}' , \lambda_I')$ share the same shift vector, $\lambda_I = \lambda_I'$, they must still label the same element of $G/H$ because they differ only by a linear generator ${\cal T} = {\cal X} - {\cal X}' = (X_{IJ} -X_{IJ}',0)$. It follows that the broken generators ${\cal X}$ constructed from \Eq{eq:affine_def} span the \textit{full space} $G/H$ of broken generators.

Second, our Adler zero constraints actually imply that the coset space $G/H$ is symmetric.  For a symmetric space {\it there exists} a basis in which the unbroken and broken generators satisfy commutation relations of the schematic form,
\eq{
{} [ {\cal T}, {\cal T} ] \sim {\cal T}, \qquad  [ {\cal T}, {\cal X} ] \sim {\cal X}, \qquad   [ {\cal X}, {\cal X} ] \sim {\cal T} .
}{eq:symm}
The first equation says that the unbroken generators form a subalgebra while the second equation says that the broken generators furnish a linear representation of the unbroken symmetry.  The third equation is the only nontrivial condition.

Remarkably, the generators we have defined in \Eq{eq:affine_def} are precisely in a basis that manifests \Eq{eq:symm} automatically.  This was not guaranteed, since any broken generator ${\cal X}=(X_{IJ},\lambda_I)$ is defined modulo addition by any unbroken generator ${\cal T}=(T_{IJ},0)$.  As it turns out, our particular broken generators are in a special basis in which they are ``mass off-diagonal'', i.e.~$X_{IJ}=0$ when $m_I^2 = m_J^2$.   While we do not have  explicit formulas for all unbroken generators, we still know they are ``mass on-diagonal'', i.e.~$T_{IJ}=0$ when $m_I^2 \neq m_J^2$ since any preserved symmetry must leave the spectrum invariant.

With this knowledge let us compute the action of the commutator $[{\cal T},{\cal X}]$ on the fields,
\eq{
	{} [{\cal T},{\cal X}] \phi_I &= [T,X]_{IJ} \phi_J  + T_{IJ} \lambda_J =  {\cal X}'\phi_I \quad \textrm{where} \quad {\cal X}'=([T,X]_{IJ}, T\lambda_I).
}{}
Since $T_{IJ}$ and $X_{IJ}$ are mass on-diagonal and off-diagonal, respectively, we know that their commutator $[T,X]_{IJ}$ is mass off-diagonal. Since only the broken generators are mass off-diagonal, this implies that the resulting generator is broken, so $[{\cal T},{\cal X}] = {\cal X}'$.  On the other hand, the commutator $[{\cal X},{\cal X}']$ acts on the fields as
\eq{
	{} [{\cal X},{\cal X}'] \phi_I &= [X,X']_{IJ} \phi_J  + X_{IJ} \lambda_J' -X_{IJ}' \lambda_J 
	 = {\cal T}'\phi_I \quad \textrm{where} \quad {\cal T}' = ([X,X^\prime]_{IJ},0).
}{}
Since $X_{IJ}$ and $X_{IJ}'$ are both mass off-diagonal and $\lambda_I$ and $\lambda_I'$ reside in the subspace of pion fields, the shift component $X_{IJ} \lambda_J' -X_{IJ}' \lambda_J$ only has support on the subspace of sigma fields.  However, we know from \Eq{eq:affine_1} that any shift component of an affine symmetry must act solely on the pion subspace, so the shift must vanish. Any symmetry without a shift is linear by definition, so we know that the commutator must produce an unbroken generator, and thus $[{\cal X},{\cal X}'] = {\cal T}'$.  

The fact that the coset space is symmetric implies that there is an automorphism of the algebra that sends ${\cal T} \rightarrow {\cal T}$ and ${\cal X} \rightarrow -{\cal X}$.  In fact, this automorphism simply flips the sign of all transformations between states $\phi_I$ and $\phi_J$ for $m_I^2 \neq m_J^2$.  Since ${\cal T}$ and ${\cal X}$ are mass on-diagonal and off-diagonal, respectively, under this automorphism they will be even and odd, respectively. The converse of our result is also known, where the Adler zero fails for theories where the coset space is not symmetric \cite{Kampf:2019mcd}.   The importance of a symmetric coset for constructing the theory of purely pions from Adler zeros was also emphasized in \cite{Low:2014nga}.

\subsubsection{Symmetry Restoration}

In summary, we have shown that perturbative unitarity together with the Adler zero condition on four- and five-particle scattering implies the existence of an underlying global symmetry. This symmetry is encoded in unbroken and broken generators that act on the fields as linear and affine field-space transformations. 
 
Given that we have derived the phenomenon of symmetry breaking, it is perhaps not so surprising that we can also derive symmetry restoration. For high-energy scattering, all dimensionful parameters become negligible relative to the momenta governing the process.  Consequently in this limit we are allowed to drop all dimensionful parameters such as masses $m_I$, cubic couplings $\alpha_{IJK}$, and shift parameters $\lambda_I$ of the affine symmetry transformation in \Eq{eq:affine_trans}.  On the other hand, the quartic couplings $\beta_{IJKL}$ and the linear components of the symmetry generators, $T_{IJ}$ and $X_{IJ}$, persist because they are dimensionless.  Hence, these dimensionless parameters encode a symmetry of the high-energy theory under which all fields transform {\it linearly}. We have also shown that $T_{IJ}$ and $X_{IJ}$ span a subspace of generators of the special orthogonal group of rotations on all the fields, as expected. This establishes our final claim: for any perturbatively unitary, Lorentz invariant, four-dimensional theory with a finite spectrum of locally interacting scalars, the Adler zero condition implies the existence of an ultraviolet symmetry that unifies the massless and massive states.

\section{Linear Sigma Model Example}
\label{sec:examples}

It will be instructive to study the implications of our results in a concrete example.    Consider a perturbatively unitary theory describing a single massive sigma field $\sigma$ and several massless pion fields $\pi_i$ for $1\leq i \leq N-1$.   In what follows, we show how the Adler zero constraints from \Eq{eq:adler4_singular}, (\ref{eq:adler4_regular}), and (\ref{eq:adler5_regular}) dictate a completely unique Lagrangian corresponding to a linear sigma model with the spontaneous symmetry breaking pattern $SO(N) \rightarrow SO(N-1)$.

Since \Eq{eq:adler4_singular} forbids pion interactions with states of equal mass, there is no cubic interaction involving an odd number of pions, so $\alpha_{ijk} = \alpha_{i\sigma\sigma} = 0$ but $\alpha_{ij \sigma}\neq0$, $\alpha_{\sigma\sigma\sigma}\neq 0$. Inserting these zeros into \Eq{eq:adler4_regular}, we obtain
\eq{
\beta_{ijkl} &= \frac{1}{m^2_\sigma}\left( \alpha_{ij\sigma}\alpha_{kl\sigma} + \alpha_{ik\sigma}\alpha_{jl\sigma} + \alpha_{il\sigma}\alpha_{jk\sigma}\right)\\
\beta_{ij\sigma\sigma} &= \frac{1}{m^2_\sigma}\left( \alpha_{ij\sigma}\alpha_{\sigma\sigma\sigma} - 2 \sum_n \alpha_{in\sigma}\alpha_{jn\sigma} \right) \\
\beta_{ijk\sigma} &= \beta_{i \sigma \sigma \sigma} = 0,
}{eq:SOeq1}
while plugging into \Eq{eq:adler5_regular} yields
\eq{
\sum_n \beta_{ijkn}\alpha_{nl\sigma} &= \beta_{ij\sigma\sigma}\alpha_{kl\sigma} + \beta_{ik\sigma\sigma}\alpha_{jl\sigma} + \beta_{jk\sigma\sigma}\alpha_{il\sigma} \\
\beta_{\sigma\sigma\sigma\sigma} \alpha_{ij\sigma} &= 3\sum_k \beta_{ik\sigma\sigma}\alpha_{jk\sigma}.
}{eq:SOeq2}
To solve these equations we go to a simplified field basis.  Since the cubic coupling in the Lagrangian is $\alpha_{ij\sigma} \pi_i \pi_j \sigma$ we can perform an $SO(N-1)$ rotation on the pion fields in order to diagonalize them, effectively setting $\alpha_{ij\sigma} = \alpha_i \delta_{ij}$
with no summation implied.

By combining the first lines of \Eq{eq:SOeq1} and \Eq{eq:SOeq2} we can eliminate $\beta_{ijkl}$ to give an equation for $\beta_{ij \sigma\sigma}$ in terms of $\alpha_{ij\sigma}$.   Plugging $\beta_{ij\sigma\sigma}$ into the second line of \Eq{eq:SOeq1} and setting $i=j$ we obtain the relation
\begin{align}
\alpha_i (3\alpha_i - \alpha_{\sigma\sigma\sigma})=0,
\end{align}
implying that either $\alpha_i = 0$ or $\alpha_i = \alpha_{\sigma\sigma\sigma}/3$.  For the former case, the pion $\pi_i$ is a free massless scalar decoupled from the rest of the theory.  We exclude this scenario without loss of generality since it corresponds to a subcase of our original setup where there is a free spectator pion.   Therefore, we assume the latter case, $\alpha_i = \alpha_{\sigma\sigma\sigma}/3$. Substituting the second line of \Eq{eq:SOeq1} into the second line of \Eq{eq:SOeq2} eliminates $\beta_{ik\sigma\sigma}$, producing the equation $\beta_{\sigma\sigma\sigma\sigma} = \alpha^2_{\sigma\sigma\sigma}/(3m^2_\sigma)$. All couplings are thus expressed in terms of $m_\sigma$, $\alpha_{\sigma\sigma\sigma}$, and various combinations of Kronecker deltas.

Plugging these couplings into \Eq{eq:L_renormalizable}, we obtain
\begin{align} \label{eq:L_adler_n}
\mathcal{L} = -\frac{1}{2} \partial^\mu \pi_i \partial_\mu \pi_i  -\frac{1}{2} \partial^\mu \sigma \partial_\mu \sigma - \frac{1}{2}m_\sigma^2 \sigma^2  -\frac{a_{\sigma\sigma\sigma}}{6}\sigma \left(\pi_i\pi_i + \sigma^2 \right) - \frac{a_{\sigma\sigma\sigma}^2}{72 m^2_\sigma}\left( \pi_i\pi_i + \sigma^2 \right)^2,
\end{align}
so the Lagrangian is fixed entirely by the mass $m_\sigma$ and cubic coupling $a_{\sigma\sigma\sigma}$.    We can also see that the Lagrangian is equivalent to that of the linear sigma model by defining a multiplet of fields $\Phi_I = (\pi_i, v+\sigma)$, so \Eq{eq:L_adler_n} becomes
\begin{align} \label{eq:L_SOn}
\mathcal{L} =  -\frac{1}{2} \partial^\mu \Phi_I \partial_\mu \Phi_I   - \frac{\lambda}{4}(\Phi_I \Phi_I - v^2)^2,
\end{align}
where the vacuum expectation value and quartic couplings are
\eq{
v = \frac{3 m_\sigma^2 }{a_{\sigma\sigma\sigma}} \qquad \textrm{and} \qquad
\lambda = \frac{a_{\sigma\sigma\sigma}^2}{18 m_\sigma^2 }.
}{}
We thus learn that there is a one-to-one mapping between the Lagrangian parameters of the ultraviolet theory in the unbroken phase and physically observable quantities in the broken phase.
The resulting theory is a linear sigma model with the spontaneous symmetry breaking pattern $SO(N) \rightarrow SO(N-1)$.

\Eq{eq:affine_def} and (\ref{eq:Tdef}) provide explicit expressions for all the generators of $SO(N)$ in the broken phase.  For the broken symmetry directions, we obtain
\eq{
	X_{i\sigma ,j}=  -X_{\sigma i ,j}=  \dfrac{\alpha_{i j \sigma}}{  m^2_\sigma} ,
}{eq:X_SO}
which are the $N-1$ matrices which realize the affine transformations.  As explained earlier, these generators are in bijective correspondence with the pion fields.  Meanwhile, commutators of these affine transformations yield
\eq{
T_{ij, kl} = X_{i\sigma,k} X_{\sigma j,l}-X_{i\sigma ,l} X_{\sigma j,k} \propto \delta_{ik}  \delta_{jl} -\delta_{il}  \delta_{jk},
}{}
which are $(N-1)(N-2)/2$ matrices which realize the linear transformations corresponding to the unbroken symmetry $SO(N-1)$.   All together, the generators $T_{ij,kl}$ and $X_{i\sigma,j}$ form the full set of $N(N-1)/2$ antisymmetric generators of the original $SO(N)$ symmetry.   

We have thus proven that the unique perturbatively unitary theory of a single massive sigma field coupled to $N-1$ massless pion fields is the $SO(N) \rightarrow SO(N-1)$ linear sigma model.  We emphasize that the {\it uniqueness} of the Lagrangian, as well as the explicit expression of {\it all} generators in terms of couplings and masses is special.  For a more general spectrum, the broken generators can always be determined but the same may not be true for the unbroken generators.  In this case it may be that only a subset of Lagrangian parameters can be fixed.  Nevertheless, we are always able to deduce the existence of an underlying symmetry that unifies the massless and massive degrees of freedom.

\section{Conclusions} \label{sec:conclusions}

Can symmetry be mandatory rather than optional?  In this paper we have argued yes, at least for a perturbatively unitary, Lorentz invariant, four-dimensional scalar theory with a finite spectrum, local interactions, and Adler zero  conditions on the massless degrees of freedom.  
 
First, we have shown that perturbative unitarity and locality enforce a strict bound on scattering amplitudes at arbitrarily high energy.  In turn, this implies that the space of theories under consideration can, upon an appropriate field redefinition, always be described by a renormalizable Lagrangian.  This is of course a familiar statement, but one derived here through scattering amplitudes.  Second, we have demonstrated that the Adler zero constraints on massless states directly imply the existence of an underlying set of unbroken and broken symmetry generators.   The latter are in bijective correspondence with the pion degrees of freedom and can be derived explicitly in terms of the couplings and masses.  Furthermore, they manifestly transform the massless and massive degrees of freedom amongst each other.  At high energies, the spectrum unifies into multiplets which linearly realize some subgroup of the special orthogonal group.

The present work leaves a number of avenues for future exploration.  For example, it would be interesting to apply our Adler zero conditions to systematically ``bootstrap'' theories starting with no input other than the spectrum.  We have already done this for the linear sigma model, but more generally the space of theories sculpted by our constraints should give a well-defined notion of the parameter space of allowed ultraviolet completions.  

Related to this is the question of whether one can constructively derive the vacuum expectation values of the ultraviolet theory directly from the affine symmetry transformations found in the broken phase.   This task is not obviously possible.  In fact, it may be impossible if there exists even a single theory exhibiting affine symmetries which do not arise purely from linear symmetries in the presence of vacuum expectation values.   It would be interesting to explore this option in the future.

Another natural direction along which to extend our results is higher spin, e.g.~with the addition of fermions.  In this case it should be possible to study Yukawa theories, as well as supersymmetric extensions like the Wess-Zumino model.  For the case of gauge bosons, the goal would be to derive the most general possible Higgs mechanism consistent with an input spectrum.  In this last scenario all pions are eaten and thus unphysical, so the Adler zero condition will be unnecessary and perturbative unitarity alone will be sufficient.   This avenue was in fact pursed long ago in a number of seminal works \cite{Lee:1977yc, Cornwall:1974km}, however with additional assumptions, e.g.~the structure of the electroweak sector or the existence of a Stueckelberg mechanism. More recently, progress has been made towards a purely on-shell description of the electroweak sector using unitarity bounds \cite{Bachu:2019ehv}. Interesting constraints on the geometry of extra-dimensional ultraviolet completions have also been derived from unitarization of higher spin scattering \cite{Bonifacio:2019ioc,Bonifacio:2020xoc}.

\begin{center} {\bf Acknowledgments}
\end{center}
\noindent 
We are grateful to Andreas Helset, James Mangan, and Julio Parra-Martinez for helpful comments on this work.
C.C. is supported by the DOE under grant no.~DE- SC0011632 and by the Walter Burke Institute for Theoretical Physics. Z.M. is supported by the NSF Graduate Research Fellowship Program under grant no.~DGE-1745301 and by the Walter Burke Institute for Theoretical Physics.

\appendix
\section{Scaling Bounds from Perturbative Unitarity} \label{app:PU}
In this appendix we derive a general bound on the high-energy behavior of tree-level scattering amplitudes.   Our general approach is similar in spirit to that of \cite{Cornwall:1974km} except we will be more explicit with our assumptions and various caveats to the argument.  As we have throughout the paper, we assume a perturbatively unitary theory with a finite spectrum of particles with interactions that are at most polynomial in derivatives.  It will also be important that all massless particles are pions which exhibit Adler zeros.

To begin, let us define some nomenclature.  For any $z$-dependent function $f(z)$, we define a ``scaling order'', $\langle f(z)\rangle =r $, to be the minimum integer exponent $r$ such that $|f(z\rightarrow \infty)| \leq q |z|^r$ for some $q>0$. The purpose of this section is to derive a $D$-dimensional generalization of the scaling bound in \Eq{eq:scaling_bound}. Here we can interpret $z$ as a deformation parameter for the kinematic invariants as defined in \Eq{eq:HE_limit}.  However, we can also think of it as {\it any} parameterization of the on-shell kinematics such that energies or momenta scale at most linearly with $z$ and where $z\rightarrow \infty$ corresponds to high-energy, fixed angle scattering\footnote{To sidestep caveats involving logarithmic running and high-energy Landau poles we assume that the large $z$ limit is such that the momenta are larger but not {\it exponentially} larger than all the masses in the spectrum.   }.

We can bound the absolute value of the four-particle scattering amplitude by
\begin{align} \label{eq:opthm}
|{\cal A}_{1,2 \rightarrow 1,2}|  \geq {\rm Im} \, {\cal A}_{1,2 \rightarrow 1,2} &= \frac{1}{2} \sum_X \int  \left( \prod_{i=3}^{N_X} \frac{d^{D-1} p_i}{(2\pi)^{D-1}}\frac{1}{2E_i} \right)(2\pi)^D \delta^D(p_1 + p_2 - p_X )  |\mathcal{A}_{1,2\rightarrow X}|^2 \\ \nonumber
&\geq  \frac{1}{2}  \int  \left( \prod_{i=3}^{N} \frac{d^{D-1} p_i}{(2\pi)^{D-1}}\frac{1}{2E_i} \right)(2\pi)^D \delta^D(p_1 + p_2 - p_X )  |\mathcal{A}_{1,2\rightarrow 3, \cdots, N}|^2.
\end{align}
In this inequality, we have used the optical theorem, which is a consequence of unitarity. The symbol $X$ labels all possible intermediate states with $N_X$ particles and total momentum $p_X$.  In the second line, we have exploited the fact that the optical theorem involves a sum of strictly positive definite terms (the exclusive cross-sections).

This inequality implies an inequality of the scaling orders of the amplitudes appearing on the left- and right-hand sides. Roughly, the amplitudes on the right cannot grow too quickly with the energy parameter $z$, or the inequality will be violated for some sufficiently large $z$, contradicting our assumption of unitarity. Strictly speaking, this inequality applies to the full amplitudes, including trees and all loop corrections.  However, the assumption of perturbativity tells us that these amplitudes must be dominated by their tree components in the $z\rightarrow \infty$ limit, so the scaling order of each amplitude must be determined by the scaling order of its tree component. This can be seen explicitly using the definition of scaling order, but the derivation is straightforward and unilluminating, so we will omit it. It follows that any inequality of amplitude scaling orders implied by the inequality \Eq{eq:opthm} is also an inequality of the scaling orders of their tree components.

In the high-energy limit, all masses can be neglected and the large $z$ dependence of all quantities is given by dimensional analysis. \Eq{eq:opthm} then implies a bound on $D$-dimensional, $N$-particle tree amplitudes,
\begin{align} \label{eq:scaling_inequality}
\langle \mathcal{A}_4 \rangle \geq (N-2)(D-2) - D + 2\langle \mathcal{A}_N \rangle.
\end{align}
Specializing to the case $N=4$, we see that $\langle \mathcal{A}_4 \rangle \leq 4-D$, where we have assumed that the four-particle amplitude has the same high-energy scaling in the forward and fixed-angle regimes.  We will return to this assumption later. Inserting the four-particle inequality back into the $N$-particle inequality we obtain
\eq{
\langle {\cal A}_N(z) \rangle \leq N + D -\frac{ND}{2}.
}{eq:genbound}
It is amusing to study this bound for various choices of dimension $D$.
For $D=2$, we find $\langle {\cal A}_N(z) \rangle \leq 2$ for all $N$, which is why two-dimensional theories can be renormalizable with arbitrarily high order interaction vertices.  In contrast, for $D=6$ the bound is $\langle {\cal A}_N(z) \rangle \leq 6-2N$, so the $N=3$ cubic vertex is the only local interaction allowed which is consistent with perturbative unitarity.  For $D> 6$ there are simply no allowed interacting theories whatsoever.  Last but not least, restricting to the case for $D=4$, we obtain
\eq{
\langle {\cal A}_N(z) \rangle \leq 4-N,
}{}
thus deriving the scaling bound given in \Eq{eq:scaling_bound}.   

Let us briefly discuss some subtleties in the above argument.  First of all, by applying the optical theorem we have assumed that the forward limit is finite.  This is not true in general, e.g.~in the presence of massless particles exchanged in the $t$-channel.  However, we have from the start assumed that all of the massless degrees of freedom are pions which exhibit Adler zeros.  Consequently, \Eq{eq:adler4_singular} implies that a pion cannot couple to two particles of equal mass, so it does not contribute via $t$-channel exchange to the forward amplitude, and there is no forward singularity at tree level.  More generally, one might worry about loop-level infrared divergences following from multi-particle exchanges of pions. In our case, it is possible to regulate any infrared divergences in these theories with physical mass regulators, e.g.~from small masses for the pions induced by explicit symmetry breaking.
 Consequently, the forward limit is finite, which implies that the inclusive cross-section appearing in the optical theorem, \Eq{eq:opthm}, is as well. The scaling order of the inclusive cross section is then determined by the finite, regulator-independent parts of the exclusive cross sections, which are represented by the expression appearing in the second line of \Eq{eq:opthm}. 
 
Second, as noted earlier, the argument above assumes that the four-particle scattering amplitude has the same scaling behavior at high energies irrespective of whether scattering is forward or fixed-angle.   This condition fails in the specific case that the quartic interaction is vanishing and there is a $t$-channel exchange of a massive particle.  In this case the left-hand side of \Eq{eq:scaling_inequality} is not $\langle \mathcal{A}_4 \rangle$, but rather zero.  In general spacetime dimensions this leads to a slightly different scaling bound than \Eq{eq:genbound} but this difference evaporates in the case of interest, $D=4$. 
Furthermore, our earlier statements about allowed interactions in $D=2$, $D=6$, and $D>6$ still hold in this case. 

\bibliographystyle{utphys-modified}
\bibliography{unification}

\end{document}